\documentclass[a4paper,11pt]{article}

    \usepackage{jheppub}            % Springer

    \usepackage[T1]{fontenc}        % font support
    \usepackage{amsmath}            % math     
    \usepackage{amsfonts}           % number sets       
    \usepackage{charter}            % text font        
    \usepackage{amssymb}            % math
    \usepackage{bm}                 % bold math
    \usepackage{braket}             % bras and kets    
    \usepackage[nottoc]{tocbibind}  % table of contents
    \usepackage{graphicx}           % plots in png
    \usepackage{pdfpages}           % plots in pdf
    \usepackage{tikz}               % diagrams
    \usepackage{tikz-cd}            % diagrams with arrows
    \usepackage[all,cmtip]{xy}      % arrows
    \usetikzlibrary{arrows.meta}    % arrows
    \usepackage{mathrsfs}           % script font
    \usepackage{setspace}           % custom line spacing
    \usepackage{subcaption}         % side by side figures
    \numberwithin{equation}{section}% section equation labels
    \DeclareMathOperator\ind{index} % index
    \DeclareMathOperator\tr{tr}     % trace
    \newcommand{\vac}{\phi_{\mathrm{vac}}}  % vacuum state

    \title{Mass Hierarchy of $\mathbb{Z}_2$ Monopoles}

    \author{Eduardo E. Quadros,}
    \author{Paulo J. Liebgott}
    \affiliation{Departamento de Física,\\ Universidade Federal de Santa Catarina,\\
    Florianópolis,\\ Brazil}

    \emailAdd{eduardoequadros@gmail.com}
    \emailAdd{paulo.liebgott@ufsc.br}

    \abstract{     
     In this work we establish every spherically symmetric non-Abelian $\mathbb{Z}_2$ monopole generated by $su(2)$ embeddings in the $SU(4)$ Yang-Mills-Higgs model minimally broken to $SO(4)$ by a symmetric second-rank tensor Higgs field. We find new monopole solutions associated with index 4 and index 10 embeddings. These solutions belong to $su(2)$ multiplets that are higher dimensional than triplets. Properties of these monopoles such as their mass and radius are calculated in the vanishing potential limit. A parallel between this result and the Standard Model hierarchy of fermion masses is considered.}

\begin{document}
    
    \maketitle

\section{Introduction}

The idea of a magnetic monopole originated as an appeal for symmetry, and since Dirac's seminal work \cite{Dirac1931} it has drawn significant attention from physicists. The monopole solutions independently discovered by 't Hooft and Polyakov \cite{tHooft1974, Polyakov1974} greatly renewed interest in the topic, and since then, it has been extensively studied in theoretical and mathematical physics. Following these early works on monopoles in non-Abelian theories, considerable effort has been devoted to understanding the structure of topological monopoles in arbitrary non-Abelian gauge groups \cite{Julia:1975ff, Bais:1977bb, Weinberg1980, Taubes:1981gw, Weinberg:1982ev}. A significant part of this interest is due to the importance of monopoles in topics such as electromagnetic duality \cite{Montonen:1977sn, GNO1977, Osborn:1979tq, Seiberg:1994pq, Seiberg:1994aj} and color confinement \cite{tHooft1975, Mandelstam:1974pi, Kronfeld:1987ri, Seiberg:1994rs, Greensite:2003bk}. In fact, the concepts of magnetic monopoles and electromagnetic duality motivated 't Hooft and Mandelstam to propose that quark confinement is a phenomenon dual to the confinement of magnetic charges in a superconductor.

Topological monopoles, such as the 't Hooft-Polyakov monopole, mainly differ from Dirac's monopole due to the fact that they are extended solutions to the field equations in non-Abelian gauge field theories. These solutions arise in Yang-Mills-Higgs theories where the gauge group $G$ is spontaneously broken to a subgroup $G_0$ by a scalar field and the vacuum manifold exhibits a nontrivial second homotopy group $\pi_2(G/G_0)$. Each monopole emerging in this context belongs to a topological equivalence class, and different equivalence classes have a one-to-one correspondence with the elements of $\pi_2(G/G_0)$. These topological classes prevent a monopole from continuously deforming into another monopole of a different topological class. Although there is a vast literature on monopoles in Yang-Mills-Higgs theories, most of it concerns cases where the scalar field is in the adjoint representation of the gauge group and, therefore, the topological sectors form the group $\mathbb{Z}$. Much less is known when the scalar field is in representations other than the adjoint one and the second homotopy group is the cyclic group $\mathbb Z_n$, in which case one obtains the so-called $\mathbb{Z}_n$ monopoles \cite{Weinberg1983, BL1988}. In particular, in $SU(n)$ Yang-Mills-Higgs theory, spontaneously broken to $SO(n)$ by a scalar field in the symmetric part of the $n\times n$ representation, one has the topological condition for $\mathbb Z_2$ monopoles \cite{KL2010, KL2013}.

 In this work, our aim is to further investigate the properties of $\mathbb Z_2$ monopoles arising from different embeddings. We consider a Yang-Mills-Higgs theory with a gauge group $SU(4)$ minimally broken to $SO(4)$ by a scalar field in the second-rank symmetric representation. We find that this framework provides enough room for the Higgs field to populate a large vacuum manifold, resulting in various kinds of monopole, all while still being tractable enough for explicit computation.  In Sec. \ref{Review} a summary of the required monopole theory is provided, and Sec. \ref{Mathematical Framework} establishes mathematical conventions and necessary formulae.. In Sec. \ref{Vacuum Decomposition} we establish how the symmetric representation transforms under such embeddings, so that the vacuum state can be decomposed providing boundary conditions for the field equations in Sec. \ref{Field Equations}. We present numerical results in Sec. \ref{Numerical Results}. Having computed the various monopoles masses we discuss their stability in Sec. \ref{Stability}. Finally we comment on how our results may relate to duality in Sec. \ref{Duality} and summarize our results in Sec. \ref{Conclusion}. The asymptotic fields in matrix form are provided in an Appendix.

%%%%%%%%%%%%%%%%%%%%%%%%%%%%%%%%%%%%%%%%%%%%%%%%%%%%%%%%%%%%%%%%%%%%%%%%%%%%%%%%%

\section{Review} \label{Review}

A magnetic monopole can be described as a vector field configuration whose curl is radially symmetric $\vec{\nabla} \times \vec{A} = (g/4\pi) r^{-2} \hat{r}, \label{rot A}$ $g$ standing for the magnetic charge. The original description of a field satisfying this is the Dirac monopole \cite{Dirac1931}
\begin{align}
    \vec{A} = \frac{g}{4\pi r}\ \frac{\hat{s}\times \hat{r}}{1-\hat{s}\cdot \hat{r}}, \label{Dirac_U1}
\end{align}
for some direction in space $\hat{s}$ and magnetic charge $g$. A striking feature of this field is the presence of an extended singularity in the semi-axis $\mathbb{R}^+\hat{s}$, the Dirac string. Requiring this term to be nonphysical ultimately leads to a charge quantization condition $eg = 2\pi \hbar n$, where $n$ is an integer number \cite{Wu:1975es}.

The next major development in the description of monopoles was achieved in models with spontaneous symmetry breaking. In the Georgi-Glashow model a scalar field transforms under the adjoint representation of the gauge group $SO(3)$. Assuming a nontrivial vacuum expectation value, $v$, results in a topologically stable smooth field solution displaying a radially symmetric curl asymptotically. Here the fields, far from the origin, assume the 't Hooft-Polyakov \cite{tHooft1974, Polyakov1974} form,
\begin{align}
    \vec{A} = \frac{\vec{T}\times \hat{r}}{er}, \ \ \
    \phi = v\ \vec{T}\cdot \hat{r}. \label{tHooft_Polyakov}
\end{align}
Where $\vec{T}$ stands for a radial direction in the algebra $su(2)$. The cross and inner products are well defined since the gauge group is three-dimensional and so is the scalar field.

It was subsequently made clear that the Abelian Dirac type monopole in $su(2)$ is gauge equivalent to t'Hooft-Polyakov solutions \cite{arafune1975topology}. This can be seen by fixing a direction in $su(2)$ and defining both fields in that direction such that they only occupy an $U(1)$ Abelian subgroup. Additionally, the string may be set as $\hat{n}=-\hat{z}$. The embedded monopole field reads, in polar coordinates,
\begin{align}
    \vec{A} = \frac{1}{er}\tan\left(\frac{\theta}{2}\right)T_3\ \hat{\varphi}, \ \ \ \phi \equiv v\ T_3. \label{Dirac}
\end{align}
Then the application of the hedgehog gauge transformation
\begin{align}
    U(\theta,\varphi)=\exp(-i\varphi T_3)\exp(-i\theta T_2)\exp(+i\varphi T_3),\label{hedgehog}
\end{align}
where $T_3$ annihilates the vacuum, returns smooth spherically symmetric fields \eqref{tHooft_Polyakov}. Equivalently the resulting fields may be rewritten as \cite{rose1995elementary}
\begin{align*}
    \vec{A} = \frac{\vec{T}\times \hat{r}}{er}, \ \ \ \phi = v\sum_{m=-1}^1Y_1^{m*}(\theta,\varphi)\ket{1,m},
\end{align*}
where $Y_1^m$ are spherical harmonics normalized to $4\pi/3$, and $\ket{1,m}$ are eigenstates of $T_3$ with eigenvalues $m=-1,0,1$.

The Lagrangian for the Yang-Mills-Higgs model with gauge symmetry group $G$ and scalar field in arbitrary representation is given by
\begin{align}L=\frac{1}{2}\tr(F_{\mu\nu}F^{\mu\nu})+\frac{1}{2}\braket{D_{\mu}\phi,D^{\mu}\phi}-\lambda V(\phi),\label{Lagrangian}
\end{align}
where $\braket{\cdot,\cdot}$ denotes the scalar product in the internal space while $\tr$ denotes the usual trace of the Lie algebra valued fields. Field strength and covariant derivative of the scalar field read
\begin{align*}
    F_{\mu\nu}&=\partial_{\mu}A_{\nu}-\partial_{\nu}A_{\mu}+ie[A_{\mu},A_{\nu}],\\
    D_{\mu}\phi&=\partial_{\mu}\phi+ieA_{\mu}\phi.
\end{align*}
while $\lambda V(\phi)$ is the scalar potential controlled by some free parameter $\lambda> 0$. The gauge group $G$ is spontaneously broken to a subgroup $G_0$ by a vacuum expectation value. This means that the scalar field is subject to a potential $V$ whose minimum $M=\{\phi:V'(\phi)=0\}$ is invariant with respect to $G_0$. This manifold of vacuum states $M$ is called the vacuum manifold and $G_0$ the unbroken gauge group. When the vacuum manifold is acted upon by $G$ transitively it becomes homeomorphic to the quotient space $M\cong G/G_0$. Solutions in this theory can be classified in topological sectors defined by elements of $\pi_2(M)\cong \pi_2(G/G_0)$ and are said to be of such type. In particular 't Hooft-Polyakov monopoles are of type $\mathbb{Z}$, since $\pi_2(SO(3)/SO(2))\cong\mathbb{Z}$. On the other hand $SU(m)\rightarrow SO(m)$ yield $\mathbb{Z}_2$ monopoles.

A solution is said to be spherically symmetric if it is invariant with respect to the generalized angular momentum $\vec{L} + \vec{T}$, where $\vec{L}=-i\hat{r}\times \vec{\nabla} $ \cite{Wilkinson97}. 't Hooft-Polyakov monopoles of unit charge, for example, are spherically symmetric. More generally, consider the asymptotic field
\begin{align*}
\phi(\theta,\varphi) = v\sum_{m=-l}^l Y_l^{m*}(\theta,\varphi)\ket{l,m}.
\end{align*}
Here $\ket{l,m}$, $m=-l,\cdots,+l$, are states of the $(2l+1)$-dimensional irreducible representation of $su(2)$ and $Y_l^m$ are spherical harmonics, $Y_l^{m*}=(-)^mY_l^{-m}$. That this field is spherically symmetric follows from $L_3 Y_l^m=m Y_l^m$, $L_{\pm}Y_l^m=\sqrt{l(l+1) - m (m \pm 1)} Y_l^m$ while similarly $T_3\ket{l,m}=m\ket{l,m}$ and $T_{\pm}\ket{l,m}=\sqrt{l(l+1) - m (m \pm 1)} \ket{l,m}$. 

When considering larger gauge groups this procedure gives rise to other  of monopole solutions. For instance, breaking $su(n)$ to $so(n)$ invariant under the Cartan automorphism, leads to solutions of topological type $\mathbb{Z}_2$  \cite{KL2010}. Consider the special case $SU(4)\rightarrow SO(4)$. To achieve this we fix a scalar field transforming under the symmetric second-rank tensor representation \cite{Li74}. The vacuum state fixed by $SO(4)$ is the diagonal state with four identical eigenvalues $(\vac)_{ij}=v\delta_{ij}/2$. The resulting vacuum manifold has homotopy group $\mathbb{Z}_2$, therefore this model supports topologically stable solutions. Consider for instance the embedding of Lie algebras $f:su(2)\hookrightarrow su(4)$, 
\begin{align}
    \frac{1}{2}\begin{pmatrix}
        z & x-iy \\
        x+iy & -z
    \end{pmatrix}
    \mapsto
    \frac{1}{2}\begin{pmatrix}
        z & x-iy & 0\  & 0\ \\
        x+iy & -z & 0\  & 0\ \\
        0 & 0 & 0\ & 0\  \\
        0 & 0 & 0\  & 0\
    \end{pmatrix}.\label{fundamental embedding}
\end{align}
Then the $\mathbb{Z}_2$ equivalent of the Abelian Dirac type monopole \eqref{Dirac} with respect to \eqref{fundamental embedding} reads
\begin{align}
    \vec{A} = \frac{1}{er}\tan\left(\frac{\theta}{2}\right)\frac{1}{2}\begin{pmatrix}
        0 & -i & 0 & 0\\
        i & 0 & 0 & 0\\
        0 & 0 & 0 & 0\\
        0 & 0 & 0 & 0
    \end{pmatrix}\hat{\varphi},\ \ 
    \vac \equiv \frac{v}{2}\begin{pmatrix}
        1 & 0 & 0 & 0\\
        0 & 1 & 0 & 0\\
        0 & 0 & 1 & 0\\
        0 & 0 & 0 & 1
        \end{pmatrix}. \label{Dirac Z2}
\end{align}

Now, the hedgehog transformation \eqref{hedgehog}, whose generators $T_i$ are given yet again by the embedding \eqref{fundamental embedding}, transforms \eqref{Dirac Z2} into an embedded fundamental $\mathbb{Z}_2$ monopole. Here, just like in the $su(2)$ model with adjoint Higgs, the scalar field is composed of a triplet of eigenstates of $T_3$. Linear combinations of mutually commuting fundamental embeddings are also a solution \cite{KL2010}, which we shall call diagonal. Fundamental and diagonal embeddings are referred to as regular, and those which are not regular are called special \cite{Lorente1972}.

%%%%%%%%%%%%%%%%%%%%%%%%%%%%%%%%%%%%%%%%%%%%%%%%%%%%%%%%%%%%%%%%%%%%%%%%%%%%%%

\section{Mathematical Framework} \label{Mathematical Framework}
Let us establish some notation and algebraic results before proceeding. The rank one semisimple Lie algebra, $su(2)$, is addressed in both the Cartesian basis $\{T_1,T_2,T_3\}$ satisfying $[T_i,T_j]=i\varepsilon_{ijk}T_k$ and the Cartan-Weyl basis $\{h,e^+,e^-\}$ where $[h,e_{\pm}]=\pm 2e_{\pm}$ and $[e^+,e^-]=h$. Let $g$ be a semisimple Lie algebra of rank $r\geq 2$. Choose $\{H^{\alpha}\}$ as basis for the Cartan subalgebra indexed by simple roots $\alpha\in \Delta$, a set of $r$ vectors in $(r+1)$-dimensional Euclidean space. Call $\Phi$ the set of all roots of $g$. In the Chevalley basis $\{H^{\alpha},E^{\beta}\}$ the bracket relations of $g$ read \cite{fuchs2003symmetries}
\begin{align}
[H^{\alpha},H^{\beta}]&=0,\label{Chevalley HH}\\
[H^{\alpha},E^{\beta}]&=\alpha\cdot\beta E^{\beta},\label{Chevalley HE} \\
[E^{\alpha},E^{\beta}]&=
    \begin{cases}
    H^{\alpha},&\mathrm{if}\ \alpha+\beta=0,\\
    N_{\alpha,\beta}E^{\alpha+\beta},&\mathrm{if}\ \alpha+\beta \in \Phi,\\
    0,&\mathrm{otherwise}.
    \end{cases}\label{Chevalley EE}
\end{align}
Every coefficient on the left hand-side is an integer; $\alpha\cdot\beta$ denotes the inner product between roots and $N_{\alpha,\beta}=\pm (p+1)$ is the largest integer, in absolute value, such that $\alpha+p\beta\in\Phi$ is still a root.

An embedding here is understood as an injective Lie algebra homomorphism, so a nontrivial linear transformation $f:su(2)\hookrightarrow su(4)$ which preserves the bracket $f([X,Y])=[f(X),f(Y)]$. We proceed to classify all $su(2)$ embeddings in $su(4)$ explicitly \cite{Lorente1972}. Let $\{h,e^+,e^-\}$ be a basis for $su(2)$. Linearity of $f$ implies each embedding is uniquely determined by its action on this basis. Moreover we require the resulting step operators to be related by antilinear Cartan involutions, i.e. $\omega (f(e^+))=-f(e^-)$, where $\omega(E^{\alpha})=-E^{-\alpha}$.

Thus we only need to determine two generators $H:=f(h)$, $E^+:=f(e^+)$ and the rest of the algebra will follow from brackets and involutions. Choose the $su(2)$ basis such that $f(h)$ belongs to the $su(4)$ Cartan subalgebra,
\begin{align}
    H=\sum_{\alpha\in\Delta}\rho_{\alpha}H^{\alpha}, \label{H}
\end{align}
for some coefficients $\rho_{\alpha}$ to be determined. Now, applying the homomorphism condition to the $su(2)$ bracket equations gives $[H,E^+]= 2E^+$ and $[E^+,E^-]=H$. From the first we see that the step operator cannot have components in the Cartan subalgebra, as those would be annihilated by \eqref{H}. Therefore it must be of the form
\begin{align}
    E^+=\sum_{\gamma\in\Phi}x_{\gamma}E^{\gamma}, \label{E}
\end{align}
for some second set of coefficients $x_{\gamma}$ to be determined as well. Equation \eqref{Chevalley HE} gives
\begin{align}   [H,E^+]&=\sum_{\gamma\in\Phi}\left(\sum_{\alpha\in\Delta}\rho_{\alpha}\alpha\right)\cdot\gamma\ x_{\gamma}E^{\gamma}. \label{HE+}
\end{align}
The linear combination in parenthesis will be called the embedding vector $\rho$. Notice that coefficients $\rho_{\alpha}$ can be recovered using the dual basis of coweights $\lambda^{\vee}$. Now the first homomorphism condition $[H,E^+]=2E^+$ is equivalent to
\begin{align}
    \rho\cdot\gamma=2, \ \ \mathrm{or}\ \  x_{\gamma}=0. \label{rho dot gamma = 2}
\end{align}
for every root $\gamma\in\Phi$. Notice that when $\gamma$ is a root, $-\gamma$ is also a root so $x_{\gamma}$ and $x_{-\gamma}$ cannot both be different from zero simultaneously. Because of this we expect $x_{\gamma}$ to be mostly zero save for a few terms. A similar computation for $[E^+,E^-]=H$ yields
\begin{align}
    \sum_{\gamma\in\Phi}|x_{\gamma}|^2\gamma=\rho, \ \ \mathrm{and}\ \  \sum_{\gamma'+\gamma''=\gamma}x_{\gamma}x^*_{\gamma'}N_{\gamma',-\gamma''}=0. \label{x^2 gamma = rho}
\end{align}
We have arrived at two sets of equations, \eqref{rho dot gamma = 2} and \eqref{x^2 gamma = rho} for two sets of variables $x_{\gamma}$ and $\rho_{\alpha}$. Multiplying the first equation in \eqref{x^2 gamma = rho} by some root $\gamma'\in\Phi$ and applying \eqref{rho dot gamma = 2} eliminates the embedding vector, so
\begin{align}
    \sum_{\gamma\in\Phi}|x_{\gamma}|^2\gamma\cdot\gamma' = 2, \qquad \mathrm{or} \qquad x_{\gamma'} = 0. \label{x^2 gamma gamma' = 2}
\end{align}
for every root $\gamma'\in\Phi$. Now for each choice of nonzero $x_{\gamma'}$ we are left with a linear system of the real variables $|x_{\gamma}|^2$. This may be solved case by case, and for each solution the embedding vector is recovered using the first equality in \eqref{x^2 gamma = rho}. We focus on $su(4)$ from now on.

There are twelve coefficients $x_{\gamma}$ and three coefficients $\rho_{\alpha}$ to be determined. Injectivity of $f$ prevents every $x_{\gamma}$ from being zero, so assume only $x_{\alpha_1}\neq 0$. In such case \eqref{x^2 gamma gamma' = 2} reduces to a single equation $|x_{\alpha_1}|^2\alpha_1\cdot\alpha_1=2$. But $|\alpha_1|^2=2$ in $su(4)$ thus $x_{\alpha_1}$ is an arbitrary phase which we fix as unity. Finally \eqref{x^2 gamma = rho} returns the embedding vector $\rho=\alpha_1$ and concludes the first solution; $\rho=\alpha_1$ and $x_{\gamma}=\delta_{\gamma\alpha_1}$. This is the fundamental embedding in the direction of the first simple root. Any other choice of root $\gamma$ satisfying $x_{\gamma}\neq 0$ would yield an embedding in the corresponding direction.

Next, suppose the two coefficients $x_{\alpha_1},x_{\alpha_3}\neq 0$. Now \eqref{x^2 gamma gamma' = 2} returns two equations, and since $\alpha_1\cdot\alpha_3=0$, they give us $|x_{\alpha_1}|^2=|x_{\alpha_3}|^2=1$, and so $x_{\gamma}=\delta_{\gamma\alpha_1}+\delta_{\gamma\alpha_3}$. Equation \eqref{x^2 gamma = rho} returns $\rho=\alpha_1 + \alpha_3$ concluding the second solution. For the case $x_{\alpha_1},x_{\alpha_2}\neq 0$ equation \eqref{x^2 gamma gamma' = 2} yields a system of equations which results in $|x_{\alpha_1}|^2=|x_{\alpha_2}|^2=2$. Thus the third solution has embedding vector $\rho=2\alpha_1+2\alpha_2$ and coefficients $x_{\gamma}=\sqrt{2}\delta_{\gamma\alpha_1}+\sqrt{2}\delta_{\gamma\alpha_2}$. Finally, for $x_{\alpha_1},x_{\alpha_2},x_{\alpha_3}\neq 0$. The system \eqref{x^2 gamma gamma' = 2} gives $|x_{\alpha_1}|^2,|x_{\alpha_2}|^2,|x_{\alpha_3}|^2=3,4,3$, so $x_{\gamma}=\sqrt{3}\delta_{\gamma\alpha_1}+2\delta_{\gamma\alpha_2}+\sqrt{3}\delta_{\gamma\alpha_3}$. The embedding vector is $\rho=3\alpha_1+4\alpha_2+3\alpha_3$. This is the fourth and final solution.

An embedding combining adjacent roots, that is $x_{\alpha_1},x_{\alpha_1+\alpha_2}\neq 0$, is prevented by the second condition in \eqref{x^2 gamma = rho}. Other selection of roots for $x_{\gamma}\neq 0$ would either be equivalent to a Weyl reflection of the ones above, or impossible, when \eqref{x^2 gamma gamma' = 2} returns an overdetermined system.

In summary there are four $su(2)$ embeddings $\{H,E^+, E^-\}$, given by the examples\cite{WILCZEK1976160}
\begin{gather}
\begin{aligned}
    H&=H^{\alpha_1},& E^+&=E^{\alpha_1};\\
    H&=H^{\alpha_1}+H^{\alpha_3},&  E^+&=E^{\alpha_1}+E^{\alpha_3};\\   
    H&=2H^{\alpha_1}+2H^{\alpha_2},&  E^+&=\sqrt{2}E^{\alpha_1}+\sqrt{2}E^{\alpha_2};\\
    H&=3H^{\alpha_1}+4H^{\alpha_2}+3H^{\alpha_3},&  E^+&=\sqrt{3}E^{\alpha_1}+2E^{\alpha_2}+\sqrt{3}E^{\alpha_3}.
\end{aligned}\label{embeddings}
\end{gather}
And $E^-=(E^+)^{\dagger}$, where $(E^{\alpha_i})^{\dagger}:=-\omega (E^{\alpha_i})=E^{-\alpha_i}$. All other solutions are gauge conjugate to the ones above.

In order to see that embeddings  are indeed distinct, we argue the following. Embeddings preserve the inner product except for a scaling factor, that is,
\begin{align}
    f(X)\cdot f(Y)=\ind(f) X\cdot Y, \label{index}
\end{align}
called the index of the embedding \cite{Lorente1972}. Note that gauge transformations preserve this inner product, so they must also preserve the index. Now, setting both $X,Y=H$ in \eqref{H}, \eqref{index} returns $\ind(f)=\rho^2/2$. We conclude that embeddings \eqref{embeddings} are organized by indices 1, 2, 4 and 10 respectively and hence distinct, so from now on they will be referred to by their index.

%%%%%%%%%%%%%%%%%%%%%%%%%%%%%%%%%%%%%%%%%%%%%%%%%%%%%%%%%%%%%%%%%%%%%%%%%%%%%%%%%%%%%%

\section{Branching Rules} \label{Branching Rules}
In the previous section we have described how to find $su(2)$ subalgebras of $su(4)$ using embeddings $f:su(2)\hookrightarrow su(4)$ expressed as \eqref{embeddings}. In order to evaluate the hedgehog gauge transformation \eqref{hedgehog} we must understand how the resulting embedded subgroup acts on the symmetric scalar field. This is done by expressing the vacuum state as a combination of $su(2)$ multiplets. We turn therefore to the task of finding those multiplets. In the symmetric second-rank tensor representation of $su(4)$, the $\mathbf{10}$, the embedded $su(2)$ no longer acts transitively on the space of states, so it decomposes this representation into a direct sum of smaller irreducible representations (irreps). Each embedding induces a distinct decomposition called its branching rule \cite{georgi2000lie}. 
For a general representation we define the basis of weight states $\{\ket{\mu}\}$ as eigenstates of the Cartan generators, $H^{\alpha}\ket{\mu}=\alpha\cdot\mu\ket{\mu}$ and for $su(4)$ they can be abbreviated
$\ket{m_1\ m_2\ m_3}=\ket{m_1\lambda_1+m_2\lambda_2+m_3\lambda_3},$
such that $H^{\alpha_i}\ket{m_1\ m_2\ m_3}=m_i\ket{m_1\ m_2\ m_3}$. The highest weight of the $\mathbf{10}$ is $\Lambda=2\lambda_1=\ket{2\ 0\ 0}$, and a diagram of the weight space is provided in Figure \ref{weight diagram}. Using the identities
\begin{align}
    E^{-\alpha}\ket{\mu}&=\sqrt{\alpha\cdot\mu}\ket{\mu-\alpha},&&
    \textrm{if\ } E^{+\alpha}\ket{\mu}=0.\label{lower once}\\
    E^{-\alpha}\ket{\mu}&=\sqrt{2\alpha\cdot\mu+|\alpha|^2}\ket{\mu-\alpha}\label{lower twice},&&\textrm{if\ }E^{+\alpha}\ket{\mu}\neq 0 \textrm{\ and\ }(E^{+\alpha})^2\ket{\mu}=0.
\end{align} it is possible to establish the branching rules of the embeddings \eqref{embeddings}. We illustrate the index 4 embedding branching rule below.

\begin{figure}
    \centering
\begin{tikzcd}[column sep=large, row sep=huge]
  \!\!\!\!\!\! \Lambda= \ket{\ 2\ 0\ 0\ } \arrow[r,"\alpha_1" ]& \ket{\ 0\ 1\ 0\ }\arrow[r,"\alpha_1" ]\arrow[d,"\alpha_2"]& \ket{-2\ 2\ 0}\arrow[d,"\alpha_2"] \\
    & \ket{1\!\ -\!1\ 1} \arrow[r,"\alpha_1"]\arrow[dl,"\alpha_3" description]& \ket{-1\ 0\ 1}\arrow[d,"\alpha_2"]\arrow[dl,"\alpha_3" description]\\
   \ket{1\ 0\!\ -\!1}\arrow[r,"\alpha_1"] & \ket{-1\ 1\!\ -\!1} \arrow[d,"\alpha_2"]& \ket{\ 0\!\ -\!2\ 2}\arrow[dl,"\alpha_3" description]\\ & \ket{0\ -\!1\ 0}\arrow[dl,"\alpha_3" description] & \\ 
\ket{0\ 0\ -\!2}& &
\end{tikzcd} 
    \caption{A diagram of the weight space for the symmetric second-rank tensor representation of $SU(4)$. The highest weight is $\Lambda = \ket{2\ 0\ 0}$ and each arrow represents a lowering operation in the direction of some root $\alpha_i$. The resulting weight is determined using \eqref{lower once} and \eqref{lower twice}. Each state $\ket{m_1\ m_2\ m_3}$ can be lowered, at most $m_i>0$ times in the $i$-th direction. For example the highest weight can be lowered twice by $\alpha_1$, the second one once by $\alpha_2$, and so forth until every path has been taken into account.}
    \label{weight diagram}
\end{figure}

Take $H=2H^{\alpha_1}+2H^{\alpha_2}$ and $E^{+}=\sqrt{2}E^{\alpha_1}+\sqrt{2}E^{\alpha_2}$. Then,
\begin{align*}
    E^-\ket{2\ 0\ 0}&=2\ \ket{0\ 1\ 0},\\
    E^-\ket{0\ 1\ 0}&=2\ \ket{-2\ 2\ 0}+\sqrt{2}\ \ket{1\ -1\ 1},\\
    E^-\left(2\ \ket{-2\ 2\ 0}+\sqrt{2}\ \ket{1\ -1\ 1}\ \right)&=6\ \ket{-1\ 0\ 1},\\
    E^- \ket{-1\ 0\ 1}&=2\ \ket{0\ -2\ 2},\\
    E^- \ket{0\ -2\ 2}&=0.
\end{align*}
Consequently the states above transform as an $su(2)$ quintuplet which we may label
\begin{align*}
    \ket{2,+2}&=\ket{2\ 0\ 0},\\
    \ket{2,+1}&=\ket{0\ 1\ 0},\\
    \ket{2,0}&=\sqrt{2/3}\ \ket{-2\ 2\ 0}+\sqrt{1/3}\ \ket{1\ -1\ 1},\\
    \ket{2,-1}&=\ket{-1\ 0\ 1},\\
    \ket{2,-2}&=\ket{0\ -2\ 2}.
\end{align*}
Now take the state orthogonal to $\ket{2,0}$ given by $\ket{0,0}'=\sqrt{1/3}\ \ket{-2\ 2\ 0}-\sqrt{2/3}\ \ket{1\ -1\ 1}$. This transforms as a singlet, i.e. $E^+\ket{0,0}'=E^-\ket{0,0}'=0$. Next, going down the diagram in Figure \ref{weight diagram}, we see $E^+\ket{1\ 0\ -1}=0$, so take it as the starting point for a new multiplet. Repeating the previous procedure we find the triplet
\begin{align*}
    \ket{1,+1}&=\ket{1\ 0\ -1},\\
    \ket{1,\ 0}&=\ket{-1\ 1\ 1},\\
    \ket{1,-1}&=\ket{0\ -1\ 0}.
\end{align*}
At last the remaining state is a second singlet: $\ket{0,0}''=\ket{0\ 0\ -2}$. Primes and double primes are appended to distinguish degenerate states. All ten states are now accounted for so the branching rule for the index $4$ embedding is completed and we summarize it as
\begin{align*}
    \mathbf{10}\xrightarrow[]{\mathrm{index\ } 4} \mathbf{5}+\mathbf{3}+\mathbf{1}+\mathbf{1}.
\end{align*}
This and all other branching rules are collected in Table \ref{branchings}. These results will be necessary for describing how the vacuum state transforms under $su(2)$ embeddings.

\begin{table}
    \centering
\begin{spacing}{1.2}
{
\begin{align}
& \ \ \ \ \mathbf{10} \xrightarrow[]{\mathrm{index\ } 1}  \mathbf{3} + \mathbf{2} + \mathbf{2} + \mathbf{1} + \mathbf{1}:\\
&\begin{cases}
        \ket{2\ 0\ 0}\longrightarrow\ket{0\ 1\ 0}\longrightarrow\ket{-2\ 2\ 0};\\
        \ket{1\ -\!1\ 1}\longrightarrow\ket{-1\ 0\ 1};\\
        \ket{1\ 0\ -1}\longrightarrow\ket{-1\ 1\ -1};\\
        \ket{0\ -2\ 2};\ \ \ \ \ket{0\ -1\ 0};\ \ \ \ket{0\ 0\ -2}.
    \end{cases}\label{branching 1}\\
& \ \ \ \ \mathbf{10} \xrightarrow[]{\mathrm{index\ } 2} \mathbf{3} + \mathbf{3} + \mathbf{3} + \mathbf{1}:\\
&\begin{cases}
        \ket{2\ 0\ 0}\longrightarrow\ket{0\ 1\ 0}\longrightarrow\ket{-2\ 2\ 0};\\
        \ket{0\ -2\ 2}\longrightarrow\ket{0\ -1\ 0}\longrightarrow\ket{0\ 0\ -2};\\
        \ket{1\ -1\ 1}\longrightarrow\left(\ket{-1\ 0\ 1}+\ket{1\ 0\ -1}\right)/\sqrt{2}\longrightarrow\ket{-1\ 1\ -1};\\
        \left(\ket{-1\ 0\ 1}-\ket{1\ 0\ -1}\right)/\sqrt{2}.
    \end{cases}\label{branching 2}\\
& \ \ \ \ \mathbf{10} \xrightarrow[]{\mathrm{index\ } 4}  \mathbf{5} + \mathbf{3} + \mathbf{1} + \mathbf{1}:\\
&\begin{cases}
        \ket{2\ 0\ 0}\longrightarrow\ket{0\ 1\ 0}\longrightarrow\left(\sqrt{2}\ \ket{-2\ 2\ 0}+\ket{1\ -1\ 1}\right)/\sqrt{3}\longrightarrow\\
        \longrightarrow\ket{-1\ 0\ 1}\longrightarrow\ket{0\ -2\ 2};\\
        \ket{1\ 0\ -1}\longrightarrow\ket{-1\ 1\ -1}\longrightarrow\ket{0\ -1\ 0};\\
        \left(\ket{-2\ 2\ 0}-\sqrt{2}\ \ket{1\ -1\ 1}\right)/\sqrt{3};\ \ \ \ket{0\ 0\ -2}.\\
    \end{cases}\label{branching 4}\\
& \ \ \ \ \mathbf{10} \xrightarrow[]{\mathrm{index\ } 10}  \mathbf{7} + \mathbf{3}:\\
&\begin{cases}
        \ket{2\ 0\ 0}\longrightarrow\ket{0\ 1\ 0}\longrightarrow\left(\sqrt{3}\ \ket{-2\ 2\ 0}+\sqrt{2}\ \ket{1\ -1\ 1}\right)/\sqrt{5}\longrightarrow \\
        \longrightarrow \left(3\ \ket{-1\ 0\ -1}+\ket{1\ 0\ -1}\right)/\sqrt{10}\longrightarrow \\
        \longrightarrow  \left(\sqrt{2}\ \ket{-1\ 1\ -1}+\sqrt{3}\ \ket{0\ -2\ 2\ }\right)/\sqrt{5}\longrightarrow\\
        \longrightarrow\ket{0\ -1\ 0}\longrightarrow \ket{0\ 0\ -2\ };\\
        \left(-\sqrt{2}\ \ket{-2\ 2\ 0}+\sqrt{3}\ \ket{1\ -1\ 1}\ \right)/\sqrt{5}\longrightarrow\\
        \longrightarrow\ \left(\ -\ket{-1\ 0\ 1}+3\ \ket{1\ 0\ -1}\ \right)/\sqrt{10}\longrightarrow\\
        \longrightarrow  \left(\sqrt{3}\ \ket{-1\ 1\ -1}-\sqrt{2}\ \ket{0\ -2\ 2\ }\right)/\sqrt{5}. \label{branching 10}
    \end{cases}
\end{align}
}
\end{spacing}
    \caption{Branching rules for the second-rank tensor symmetric representation of $su(4)$ with respect to the $su(2)\hookrightarrow su(4)$ embeddings listed in \eqref{embeddings}. Here the highest weight is successively lowered until the first multiplet is completed. Next, the remaining multiplets are found by either searching for highest weight candidates in Diagram \ref{weight diagram} or by taking a known zero weight state and writing its orthogonal complement. This procedure is repeated until dimensions on both sides match.}
    \label{branchings}
\end{table}

\section{Vacuum Decomposition} \label{Vacuum Decomposition}

In this section we aim to determine the vacuum decomposition into multiplets of the embedding in $T_3$ diagonal basis, where $T_3$ is the generator of the unbroken group. So far we have dealt with general $su(2)$ embeddings. Now we would like to narrow down our search by setting the unbroken subalgebra as $so(4)$ and requiring that exactly one $su(2)$ generator belongs to it. We do this by fixing a vacuum state $\vac$ and verifying that its unbroken gauge subalgebra is indeed $so(4)$. 

Let $\{\ket{e_1},\ket{e_2},\ket{e_3},\ket{e_4}\}$ be the canonical basis of the fundamental representation, $\mathbf{4}$. An induced basis for the symmetric second-rank tensor representation $\mathbf{10}=\mathbf{4}\otimes_s\mathbf{4}$, is written as symmetric combinations of $e_{ij}=\ket{e_i}\otimes\ket{e_j}$, where $e_{ij}$ is the canonical matrix satisfying $(e_{ij})^{kl}=\delta_i^k\delta_j^l$. A gauge transformation, also in matrix representation, $D(g)$ acts on such states like $g\cdot S = D(g)S D(g)^T$. Therefore, by choosing the vacuum as a multiple of the identity, as in \cite{KL2010},
\begin{align}
\vac= \frac{v}{4}\sum_{i=0}^4 e_{ii},\label{vacuum canonical}
\end{align}
$v$ is its expectation value, we see that this is invariant under orthogonal transformations. Thus the unbroken gauge algebra is $so(4)$.
The results of the previous section are given in Dynkin basis however, so we turn to the task of writing the vacuum state in the same way. 

To do this, note that the highest weight of the fundamental representation is $\ket{1\ 0\ 0}= \ket{e_1}$. And so, the highest weight of the $\mathbf{10}$ is $\ket{2\ 0\ 0}=\ket{e_1}\otimes\ket{e_1}=e_{11}$. Generators act on states as infinitesimal transformations, $E^{\alpha}S=D(E^{\alpha})S+SD(E^{\alpha})^T$ and 
in particular simple roots satisfy $D(E^{\alpha_i})=e_{i,i+1}$.
Now, by applying the first lowering operator $D(E^{-\alpha_1})=e_{21}$
to the highest weight we have, on one hand, $E^{-\alpha_1}\ket{2\ 0\ 0}=\sqrt{2}\ \ket{0\ 1\ 0}$, while on the other, $E^{-\alpha_1}e_{11}=e_{21}e_{11}+e_{11}e_{12}=e_{12}+e_{21}$, thus $e_{12}+e_{21}=\sqrt{2}\ \ket{0\ 1\ 0}$. Similarly, applying $E^{-\alpha_1}$ to both sides of this result, we find $e_{22}=\ket{-2\ 2\ 0}$.

The seven remaining states can be determined recursively, starting from a known state and lowering it by the appropriate simple root labeled in Figure \ref{weight diagram}. The vacuum state \eqref{vacuum canonical} now reads, in Dynkin basis,
\begin{align}
    \vac=\frac{v}{4}(\ket{2\ 0\ 0}+\ket{-2\ 2\ 0}+\ket{0\ -2\ 2}+\ket{0\ 0\ -2}).\label{vacuum dynkin}
\end{align}
This allows us to make use of the results of Sec. \ref{Branching Rules} and to determine how the vacuum transforms under the action of general $su(2)$ embeddings.

Now the aim is to specify how to write such vacuum state in the basis of the unbroken gauge algebra generator. From this we will conclude that the vacuum transforms under the embedded $su(2)$ as a linear combination of several multiplets simultaneously and in different proportions. These values will be necessary in order to set the boundary conditions for the field equations.

From embeddings \eqref{embeddings} we define generators
\begin{align}
    T_1=\frac{1}{2}H,\ \ \ \ T_2=\frac{1}{2}(E^++E^-),\ \ \ \ T_3=\frac{1}{2i}(E^+-E^-).\label{generators}
\end{align}
Note that results of Sec. \ref{Branching Rules} are already in $T_1$ diagonal basis, so we start writing the vacuum state in general terms
\begin{align}
    \vac=\sum_{b,m} v_{bm}\ket{l_b,m}_{T_1},\label{vacuum T1}
\end{align}

Where $\ket{l,m}_{T_1}$ are states of the $(2l+1)$-dimensional irrep of $su(2)$, summing over the allowed values of $m$ for each irrep, labelled by $b$, and $v_{bm}$ are coefficients to be computed.  We append a `$T_1$' in order to distinguish these states from the subsequent $T_3$ diagonal basis. Now, by referring to Table \ref{Branching Rules} and looking for the four components of the vacuum \eqref{vacuum dynkin} among the multiplets we see that \eqref{vacuum T1} reads, for the first two cases,
\begin{align}
\vac&=\frac{v}{2}(\ket{1,+1}_{T_1}+\ket{1,-1}_{T_1}+\ket{0,0}'_{T_1}+\ket{0,0}''_{T_1}),&\ind(f)&=1, \\
&=\frac{v}{2}(\ket{1,+1}'_{T_1}+\ket{1,-1}'_{T_1}+\ket{1,+1}''_{T_1}+\ket{1,-1}''_{T_1}),&\ind(f)&=2,
\end{align}
Index $4$ and $10$ embeddings require elimination of mixed states. For instance, index $4$ branching rules in Table \ref{Branching Rules} give
\begin{align*}
    \ket{2,0}_{T_1}=\sqrt{2/3}\ket{-2\ 2\ 0}+\sqrt{1/3}\ket{1\ -1\ 1},\\
    \ket{0,0}_{T_1}'=\sqrt{1/3}\ket{-2\ 2\ 0}-\sqrt{2/3}\ket{1\ -1\ 1}.
\end{align*}
After eliminating the undesired state we see that 
\begin{align*}
\ket{-2\ 2\ 0}=\sqrt{2/3}\ \ket{2,0}_{T_1}+\sqrt{1/3}\ \ket{0,0}_{T_1}'.
\end{align*}
Thus, in total, the index $4$ vacuum decomposition is
\begin{align}
    \vac&=\frac{v}{2}(\ket{2,+2}_{T_1}+\sqrt{2/3}\ket{2,0}_{T_1}+\sqrt{1/3}\ket{0,0}_{T_1}'+\ket{2,-2}_{T_1}+\ket{0,0}_{T_1}'').
    \label{decomposition T1 index 4}
\end{align}
A similar computation for the index $10$ vacuum decomposition gives
\begin{align*}
    \vac=\frac{v}{2}(&\ket{3,+3}_{T_1}+\sqrt{3/5}\ket{3,+1}_{T_1}-\sqrt{2/5}\ket{1,+1}_{T_1} + \\
    +&\ket{3,-3}_{T_1}+\sqrt{3/5}\ket{3,-1}_{T_1}-\sqrt{2/5}\ket{1,-1}_{T_1}).
\end{align*}

Now we would like to rewrite these results in terms of $T_3$ eigenstates. Note that $D(T_3)\in so(4)$ so it annihilates the vacuum state. Consequently, in $T_3$ diagonal basis the vacuum state must decompose like $\vac=\sum_b\ v_b\ket{l_b,0}$, where $\ket{l,m}$ are eigenstates of $T_3$, summed over multiplets for some coefficients $v_b$ to be determined. The change of basis from $T_1$ to $T_3$ is orthogonal, namely $\exp(i\pi T_2/2)$, and preserves multiplets, since it is generated by the embedding. Therefore the values $v_b$ must satisfy $v_b^2=\sum_mv_{bm}^2$, where $v_{bm}$ are the coefficients of \eqref{vacuum T1} we have just computed. For example, from \eqref{decomposition T1 index 4}, the index 4 case yields two terms; a quintuplet with squared coefficient $v_1^2=v^2(1+2/3+1)/4=2v^2/3$, and a singlet satisfying $v_2^2=v^2(1/3 + 1)/4=v^2/3$.

Finally the decomposition of the vacuum state into multiplets of each embedding in $T_3$ diagonal basis is given by
\begin{gather}
\begin{aligned}
\vac/v&=\sqrt{1/2}\ket{1,0}+\sqrt{1/2}\ket{0,0},&\ind(f)&=1, \\
        &=\sqrt{1/2}\ket{1,0}'\!+\sqrt{1/2}\ket{1,0}'',&\ind(f)&=2, \\
        &=\sqrt{2/3}\ket{2,0}+\sqrt{1/3}\ket{0,0},&\ind(f)&=4, \\
        &=\sqrt{4/5}\ket{3,0}+\sqrt{1/5}\ket{1,0} ,&\ind(f)&=10. 
\end{aligned}\label{scalar v_b}
\end{gather}
Where, once again, $\ket{l_b,m}$ are eigenstates of $T_3$, so for $b=1,2$,
\begin{align*}
    T_3\ket{l_b,m}=m\ket{l_b,m},\ \ \ \  \vec{T}^2\ket{l_b,m} = l_b(l_b+1)\ket{l_b,m}.
\end{align*}
Notice how the vacuum state displays a different structure for indices 4 and 10, since part of it lies in a quintuplet and a septuplet respectively. With these results now available we turn to the task of calculating the asymptotic Higgs field. This will provide boundary conditions for the field equations of the embedded monopoles.

%%%%%%%%%%%%%%%%%%%%%%%%%%%%%%%%%%%%%%%%%%%%%%%%%%%%%%%%%%%%%%%%%%%%%%%%%%%

\section{Field Equations} \label{Field Equations}

We are now set to determine the monopole fields. To do this we first compute the asymptotic fields, then calculate the radial field equations and finally solve those numerically with boundary conditions set by their asymptotic expressions.

Let $f:su(2)\hookrightarrow su(4)$ be the embeddings $f(h)=H$, $f(e^+)=E^+$ as described in \eqref{embeddings}. The third generator, $T_3$, must belong to the Lie algebra of the unbroken group $G_0$, $T_3\vac=0$, so that the Dirac type monopole configuration
\begin{align}
    \vec{A}=\frac{1}{er}\tan\left(\frac{\theta}{2}\right)T_3\ \hat\varphi,\ \ \ \phi =\sum_{b=1}^2 v_b\ket{l_b,0},\label{Abelian Dirac}
\end{align}
is Abelian, $v_b$ are the respective multiplet coefficients in \eqref{scalar v_b} satisfying $v^2=v_1^2+v_2^2$. Because the vacuum has only $m=0$ components the hedgehog local gauge transformation
\begin{align*}
    U(\theta,\varphi)=\exp(-i\varphi T_3)\exp(-i\theta T_2)\exp(+i\varphi T_3),
\end{align*}
turns \eqref{Abelian Dirac} into a spherically symmetric field configuration 
\begin{align}
    \vec{A} = \frac{\vec{T}\times \hat{r}}{er}, \ \ \ \phi = \sum_{b=1}^2 v_b\sum_{m=-l_b}^{l_b} Y_{l_b}^{m*}(\theta,\varphi)\ket{l_b,m}.\label{asymptotic field}
\end{align}
Here $Y_{l_b}^m$ are spherical harmonics satisfying the normalization convention \begin{align}
\int_0^{2\pi}\!\! \int_0^\pi Y_l^{m*}(\theta,\varphi) Y_{l'}^{m'}(\theta,\varphi)\sin\theta d\theta d\varphi = \frac{4\pi}{2l+1} \delta_{ll'}\delta^{mm'}. \label{harmonic norm}
\end{align} while $\ket{l_b,m}$ are the $T_3$-eigenstates \eqref{branching 1}-\eqref{branching 10}.

In the previous section we have determined spherically symmetric solutions in the asymptotic sphere \eqref{asymptotic field}. Now to establish the field solution everywhere we modulate each field by some radial profile functions $K, H_1, H_2$ of the dimensionless variable $\xi=ver$, i.e.
\begin{align}
    \vec{A}=v^2 e\frac{1-K(\xi)}{\xi^2}\ \hat{r}\times \vec{T},
    \ \ \ \phi=\sum_{b=1}^2\frac{H_b(\xi)}{\xi}\sum_{m=-l_b}^{l_b}\ Y_{l_b}^{m*}(\theta,\varphi)\ket{l_b, m}.\label{radial fields}
\end{align} 
Where $K(\xi)$ and $H_b(\xi)$ are radial profile functions to be determined; $Y_{l_b}^m$ are spherical harmonics normalized as in \eqref{harmonic norm} and $\ket{l_b, m}$, $b=1,2$, $m=-l_b,...,l_b$, eigenstates of $T_3$ as defined in \eqref{generators}.

The Lagrangian of our model is
\begin{align*}
L=\frac{1}{2}\tr(F_{\mu\nu}F^{\mu\nu})+\frac{1}{2}\braket{D_{\mu}\phi,D^{\mu}\phi}-\lambda V(\phi). 
\end{align*}Substituting field expressions \eqref{radial fields} into the dimensionless Hamiltonian radial density $\mathscr{H}=-(e/ 4\pi v)\int L\ d^2\Omega$, we find
\begin{align}
\mathscr{H}&=\frac{\rho^2}{2} K'^2+\frac{\rho^2}{4\xi^2}(K^2-1)^2+\frac{1}{2\xi^2}\sum_{b=1}^2\left[(\xi H_b'-H_b)^2+l_b(l_b+1)K^2H_b^2\right]+\lambda V(\phi).\label{radial hamiltonian}\end{align}
Coefficients such as the embedding index, $\rho^2/2$, and the Casimir eigenvalues, $l_b(l_b+1)$, originate from identities $\tr (T^iT^j)=\delta^{ij}\rho^2/4$ and $\bra{l,m}T^iT^i\ket{l',m'}=\delta_{ll'}\delta^{mm'}l(l+1)$ respectively. Orthogonality of the spherical harmonics and orthogonality of states are used to compute the surface integral. 

We restrict our attention to the vanishing potential limit, by taking $\lambda\rightarrow 0$. In this limit masses are minimal and the numerical method is most straightforward. Minimizing \eqref{radial hamiltonian} as a functional of the profile functions yields the second order system of three ordinary differential equations
\begin{align}
\xi^2 K'' &=  \sum_{b=1}^2\frac{l_b(l_b+1)}{\rho^2}KH_b^2 + K(K^2-1), \label{field equations K}\\
\xi^2 H_b''&= l_b(l_b+1)K^2 H_b.\label{field equations H}
\end{align}subject to the boundary conditions
\begin{align}
    K(0)=1,\ \ H_b(0)=0,\ \ K(\xi)\rightarrow 0,\ \ H_b(\xi)\rightarrow (v_b/v)\xi- w_b,\ \mathrm{\ as\ }\ \xi \rightarrow \infty. \label{limit}
\end{align}Here $v_b$ is the vacuum expectation value of the $b$-th multiplet \eqref{scalar v_b} and $w_b$ are tail parameters to be determined numerically. These parameters are only nonzero in the vanishing potential limit, where the Higgs interaction range tends to infinity. In the original BPS solution this tail parameter is one. 

Substituting the solutions of the field equations \eqref{field equations K}-\eqref{field equations H} back into the Hamiltonian density \eqref{radial hamiltonian} and integrating over $\xi$ returns the mass, in units of $M_0=4\pi v/e$. The radius of the core is defined as the critical point of the Hamiltonian density. We are ready to proceed and calculate these properties numerically.

%%%%%%%%%%%%%%%%%%%%%%%%%%%%%%%%%%%%%%%%%%%%%%%%%%%%%%%%%%%%%%%%%%%%%%%%%%%%%%

\begin{figure}[htbp]
    \centering
    \begin{subfigure}{0.45\textwidth}
        \centering
        \includegraphics[width=\linewidth]{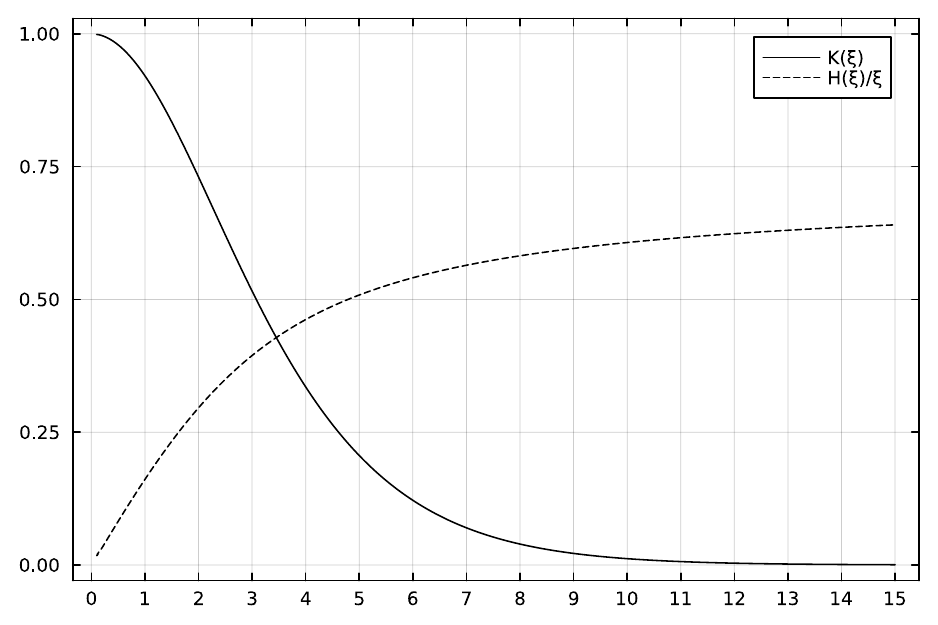}
        \caption{Index 1 Radial Fields}
    \end{subfigure}
    \begin{subfigure}{0.45\textwidth}
        \centering
        \includegraphics[width=\linewidth]{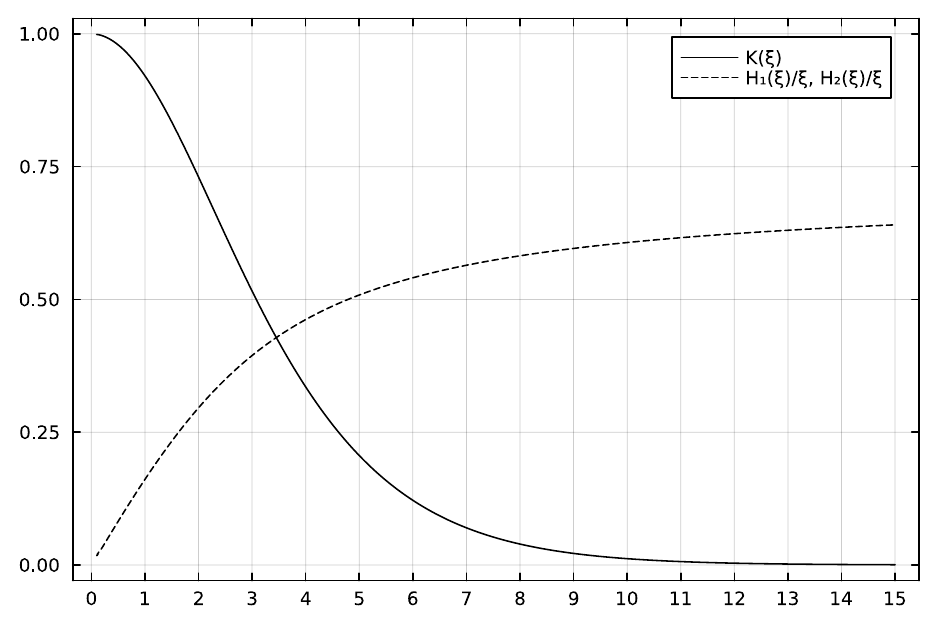}
        \caption{Index 2 Radial Fields}
    \end{subfigure}
    \\
    \begin{subfigure}{0.45\textwidth}
        \centering
        \includegraphics[width=\linewidth]{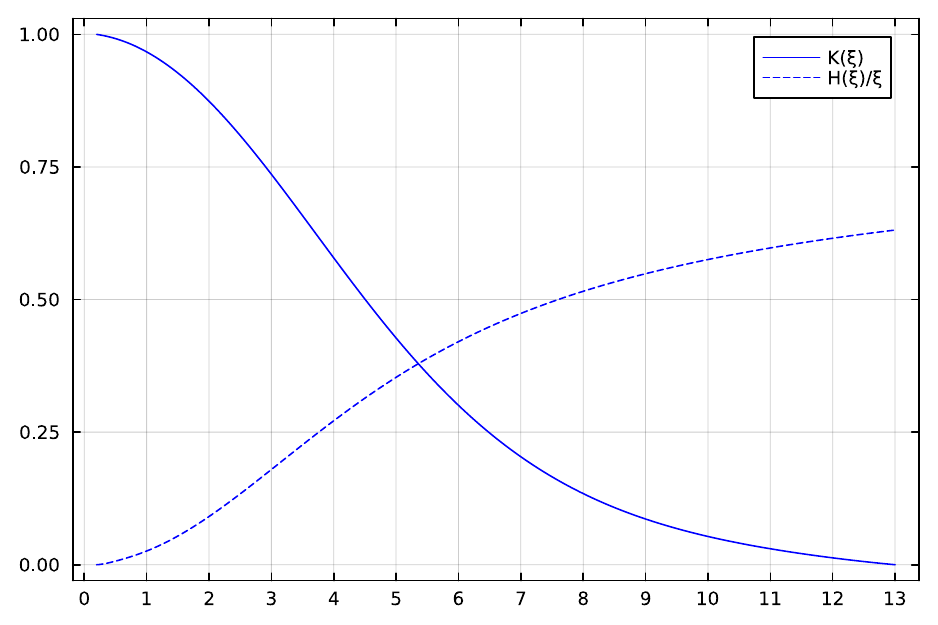}
        \caption{Index 4 Radial Fields}
    \end{subfigure}
    \begin{subfigure}{0.45\textwidth}
        \centering
        \includegraphics[width=\linewidth]{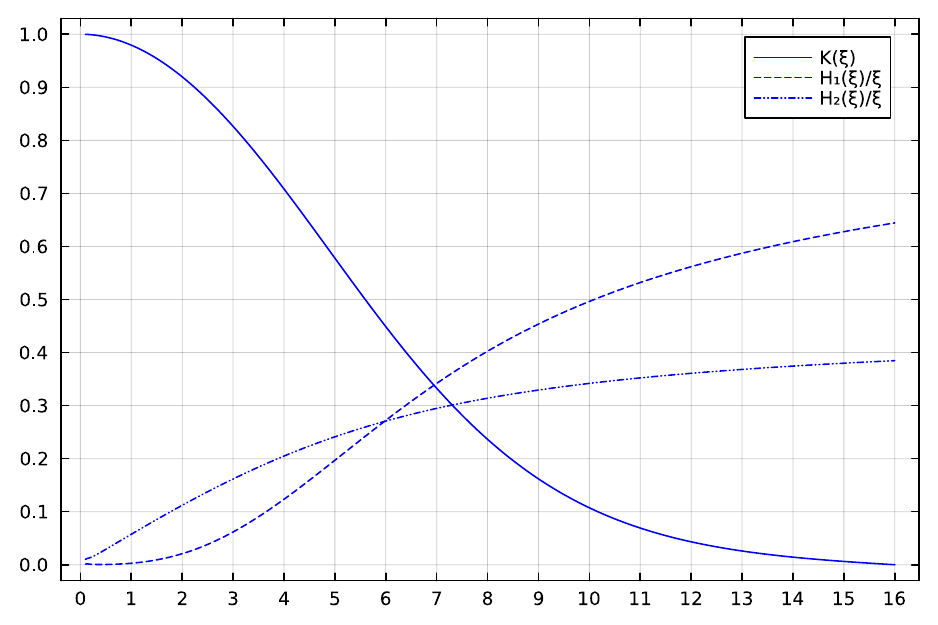}
        \caption{Index 10 Radial Fields}
    \end{subfigure}
    \caption{Profile functions of every $\mathbb{Z}_2$ monopole as functions of the dimensionless radial variable $\xi=ver$ in the vanishing potential limit $\lambda\rightarrow0$. Decaying gauge potential $K(\xi)$ is drawn in solid lines. Types of rising scalar profiles differ depending on the branching rules of the $su(2)$ embedding. (a) Index 1, $H(\xi)/\xi$ of the triplet dashed; (b) Index 2, two triplets $H_1(\xi)/\xi$ and $H_2(\xi)/\xi$ dashed are identical; (c) Index 4, quintuplet $H(\xi)/\xi$ dashed; (d) Index 10, quintuplet $H_1(\xi)/\xi$ dashed and triplet $H_2(\xi)/\xi$ dash-dotted. Singlets are always omitted.}\label{fields plot}
\end{figure}

\begin{figure}[htbp]
    \centering
    \begin{subfigure}{0.45\textwidth}
        \centering
        \includegraphics[width=\linewidth]{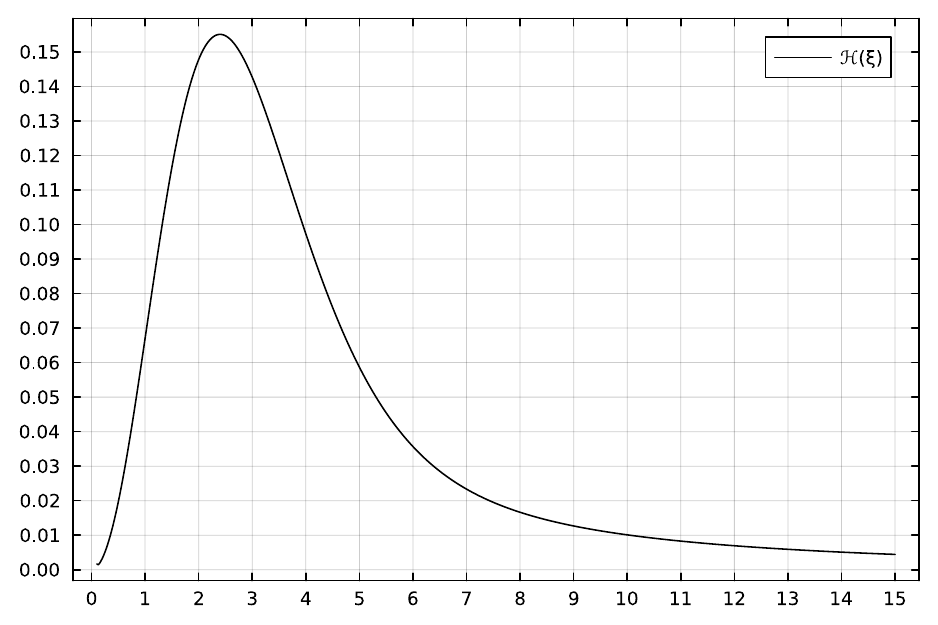}
        \caption{Index 1 Energy Density}
    \end{subfigure}
    \begin{subfigure}{0.45\textwidth}
        \centering
        \includegraphics[width=\linewidth]{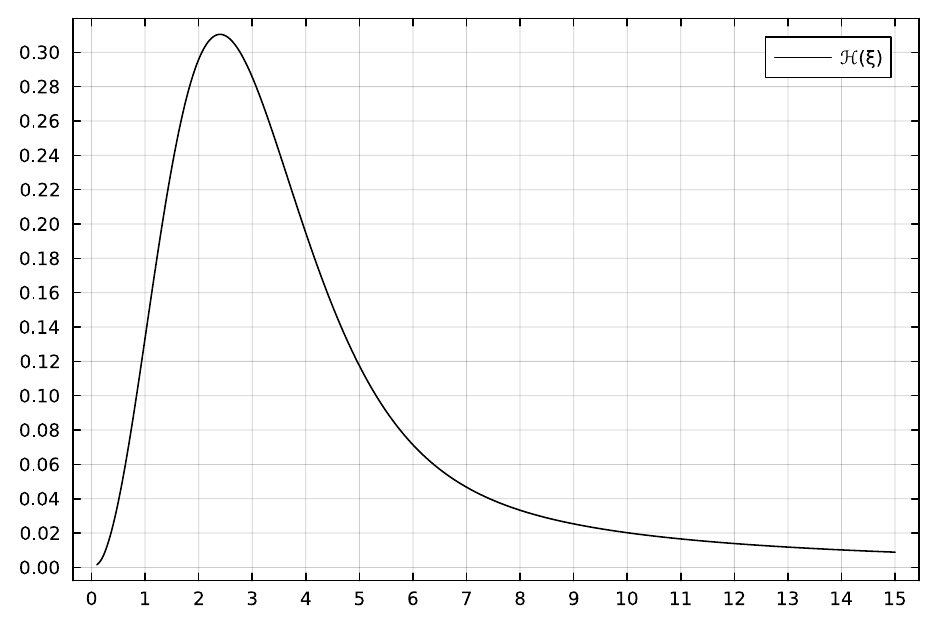}
        \caption{Index 2 Energy Density}
    \end{subfigure}
    \\
    \begin{subfigure}{0.45\textwidth}
        \centering
        \includegraphics[width=\linewidth]{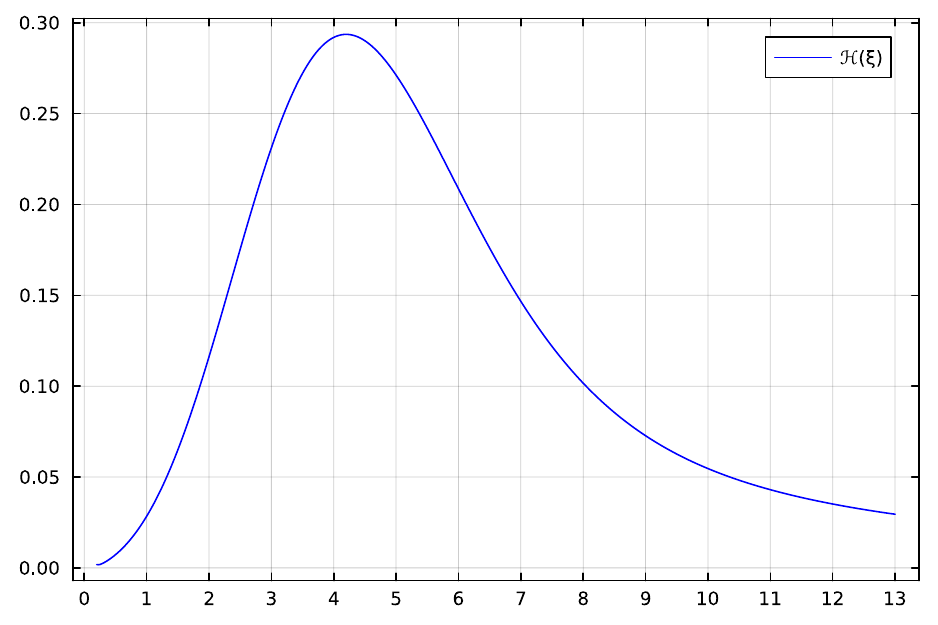}
        \caption{Index 4 Energy Density}
    \end{subfigure}
    \begin{subfigure}{0.45\textwidth}
        \centering
        \includegraphics[width=\linewidth]{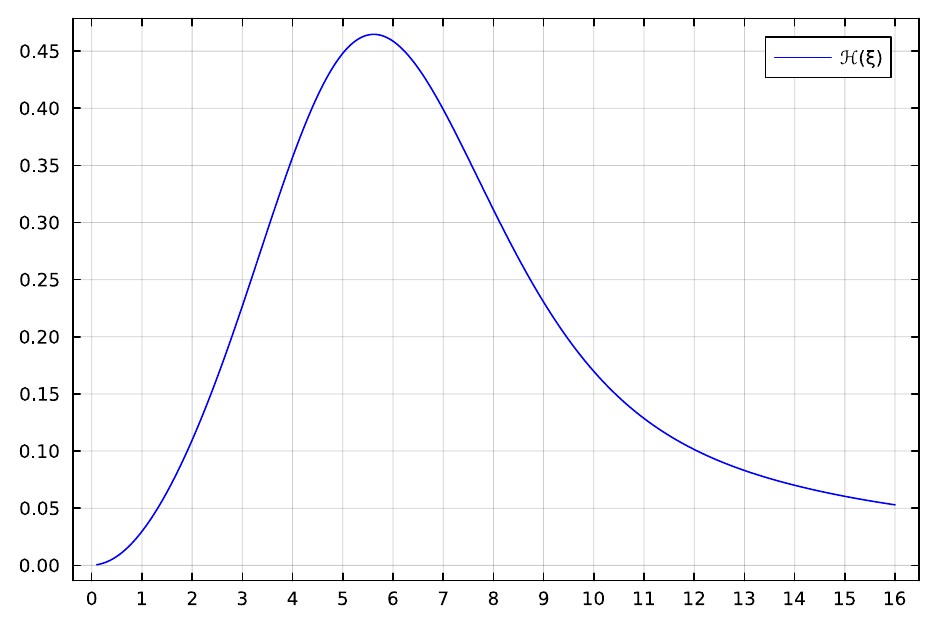}
        \caption{Index 10 Energy Density}
    \end{subfigure}
    \caption{Hamiltonian radial densities of $\mathbb{Z}_2$ monopoles of indices 1, 2, 4 and 10, functions $\mathscr{H}(\xi)$ of the dimensionless variable $\xi=ver$ in the vanishing potential limit. Their respective integrals yield the particles mass while the critical points define their radii. These values are gathered in Table \ref{results}.}
    \label{energy plot}
\end{figure}

%%%%%%%%%%%%%%%%%%%%%%%%%%%%%%%%%%%%%%%%%%%%%%%%%%%%%%%%%%%%%%%%%%%%%%%%%%%%%

\section{Numerical Results} \label{Numerical Results}
We turn to the task of computing the field configurations of $\mathbb{Z}_2$ monopoles. Given that $\mathbb{Z}_2$ monopoles of indices 4 and 10 do not belong to an $su(2)$ triplet, the BPS factorization cannot be applied here \cite{BPS1976}. Therefore, the system of ordinary differential equations \eqref{field equations K}-\eqref{limit}, in the limit $\lambda\rightarrow 0$, is solved numerically using a Runge-Kutta method in julia programming language \cite{Julia-2017,DifferentialEquations.jl-2017,CASH1996165}. 

Notice how, since we do not know $w_b$ yet, the $H_b$ limit in \eqref{limit} is not properly defined. To find this value start with a guess $w_b^0$ and read the derivative of the numerical solution at the end point, $H_b'(\xi_{\mathrm{max}})$. This provides a new value $w_b^{\mathrm{new}}= H'_b(\xi_{\mathrm{max}})\xi_{\mathrm{max}}-H_b(\xi_{\mathrm{max}})$. A few iterations of this search converges to the constants shown in Table \ref{results}. It may be worth mentioning that boundaries $\xi_{\mathrm{min}}<\xi<\xi_{\mathrm{max}}$ have been deemed reasonable when $\xi^2\mathscr{H}$ is close to constant on both ends. Because of this the tail contribution to the mass is well approximated by $\Delta M_{\mathrm{max}}/M_0= \int_{\xi_{\mathrm{max}}}^{\infty}\mathscr{H}d\xi=\mathscr{H}(\xi_{\mathrm{max}}){\xi^{-1}_{\mathrm{max}}}$. Similarly $\Delta M_{\mathrm{min}}/M_0=\mathscr{H}(\xi_{\mathrm{min}}){\xi_{\mathrm{min}}}$.
These contributions to the mass are added after-the-fact.

With these considerations in mind we compute the radial fields and energy densities which are provided in Figures \ref{fields plot} and \ref{energy plot} respectively. Each resulting mass is given in units of $M_0=4\pi v/e$ and radius in units of $R_0=(ve)^{-1}$. Solutions of indices 1 and 2 are scaled versions of the exact BPS solution \cite{Prasad:1975kr}, i.e. $H_b(\xi)=H_{\mathrm{BPS}}(\sqrt{2}\xi)$ and $K(\xi)=K_{\mathrm{BPS}}(\sqrt{2}\xi)$, yielding masses $M_1=(\sqrt{2}/2) M_0$ and $M_2=\sqrt{2} M_0$ respectively. We compute these to make sure that the algorithm used is sound. In this case, numerical solutions reflect the expected values with a relative error of $0.01\%$. Special $\mathbb{Z}_2$ monopoles of indices 4 and 10 on the other hand are novel. They return masses $M_4=2.001 M_0$ and $M_{10}=4.057 M_0$ respectively. Their radii are also larger; measuring $R_4=4.2R_0$ and $R_{10}=5.6R_0$. These and other relevant properties are organized in Table \ref{results}.
\\
\begin{table}
\begin{spacing}{1.2}
    \centering
 \begin{tabular}{ |c|c|c|c|c|c|c| } 
     \hline
     Index & Mass & Radius& $\xi^2 \mathscr{H}$ & $w_1$ & $w_2$ \\ 
     \hline
        $1$ & $0.707$ & $2.4$ & $1.000$ & $1.00$ & $0.00$\\
        $2$ & $1.414$ & $2.4$ & $2.000$ & $1.00$ & $1.00$\\
        $4$ & $2.001$ & $4.2$ & $4.993$ & $2.41$ & $0.00$\\
        $10$ & $4.057$ & $5.6$ & $13.585$ & $1.14$ & $3.96$ \\
     \hline
    \end{tabular}
    \end{spacing}
    \caption{Monopole properties organized by the indices of their respective embeddings. These results are calculated in the vanishing potential limit, $\lambda\rightarrow 0$, in which masses are minimal. Each mass is given in units of $M_0=4\pi v/e$ and radius in units of $R_0=(ve)^{-1}$. The surface energy $\xi^2 \mathscr{H}$ converges to a constant in the $\xi\rightarrow +\infty$ limit. Likewise for the scalar tail constants defined by $\xi\phi(0,0,z)\rightarrow \xi\vac+w_{b}\ket{l_b,0}$.}
    \label{results}
\end{table}

%%%%%%%%%%%%%%%%%%%%%%%%%%%%%%%%%%%%%%%%%%%%%%%%%%%%%%%%%%%%%%%%%%%%%%%%%%%%%%
\section{Stability}\label{Stability}

The model we are considering has a vacuum manifold of second homotopy type $\mathbb{Z}_2$. This implies the existence of only two topologically distinct solutions. These can be characterized by the magnetic weight of the solution. The magnetic weight is the linear combination of coweights of the unbroken algebra that defines the $T_3$ generator, namely,
\begin{align*}
    T_3=\sum_{\omega\in\Delta_0} \beta_\omega H^{\omega}
\end{align*}
where $\Delta_0=\{\omega_1,\omega_2\}$ denotes a set of simple roots and $H^{\omega}$ are Cartan generators for $so(4)$. Having this basis fixed, and by projecting roots of $su(4)$ onto roots of $so(4)$, embeddings \eqref{embeddings} of indices 1, 2, 4 and 10 yield magnetic weights $\beta=\lambda_1$, $\lambda_1+\lambda_2$, $2\lambda_1$ and $3\lambda_1+\lambda_2$, respectively, where $\lambda_i$ are weights of $so(4)$. Note that $\lambda_i^{\vee}=\lambda_i$ in this case. Monopoles of magnetic weights which belong to the root lattice $\Lambda(so(4)^{\vee})=\{n_1\omega_1+n_2\omega_2\ |\ n_1,n_2\in\mathbb{Z}\}$ or $\lambda_1+\lambda_2+\Lambda(so(4)^{\vee})$ are in the trivial sector of $\mathbb{Z}_2$, whereas monopoles of magnetic weights in the lattice $\lambda_1+\Lambda(so(4)^{\vee})$ or $\lambda_2+\Lambda(so(4)^{\vee})$ are in the nontrivial sector.

Since $\omega_i=2\lambda_i$, the integer combination $\sum_i n_i\lambda_i$ belongs to the trivial sector whenever $\sum_in_i$ is even, otherwise it belongs to the nontrivial sector. Furthermore, given that an arbitrary sum of integers is even if and only if the sum of their squares is even, we conclude that the topological class of $\mathbb{Z}_2$ monopoles is given by the parity of their index.

From the previous discussion we gather that the index 1 $\mathbb{Z}_2$ monopole is in the nontrivial sector whereas index 2, 4 and 10 monopoles are all on the same topological sector as the vacuum. Now the question of wether they are unstable or metastable remains. That is to say, if they are in a saddle point or in a false vacuum of the energy functional. This can be addressed by introducing linear perturbations to the monopole fields and verifying if they remain small or amplify over time.

Following the prescription provided in \cite{DEGLMANN2020135476}, we are able to assess the stability of $\mathbb{Z}_2$ monopoles. Note that the same line of reasoning is applicable to other field representations. At first, following their construction we conclude that perturbations to the scalar field do not introduce instabilities. Perturbations of the gauge field on the other hand may lead to instabilities if the variation belongs to the unbroken algebra. Embedded monopoles of generator $T_3$ display an unstable mode in the direction $X\in so(4)$ of the unbroken algebra if and only if
\begin{align}
[T_3,X]=qX, \mathrm{\ for\ some\ } |q|\geq 1.\label{unstable mode}
\end{align} 
This condition is rewritten in the space of $so(4)$ generators, $M_{ij}=-i(e_{ij}-e_{ji})$, as the six-by-six matrix of the adjoint action of the $T_3$ generator. Substituting $T_3 = (E^+-E^-)/2i$ as in \eqref{embeddings} into the unstable mode condition \eqref{unstable mode}, the eigenvalues $q$ for indices 1, 2, 4 and 10 yield $\{0,0,\pm1/2,\pm 1/2\}$, $\{0,0,0,0,\pm1\}$, $\{0,0,\pm1,\pm 1\}$, and $\{0,0,\pm1,\pm 2\}$ respectively.
Therefore we come to the conclusion that only the index 1 $\mathbb{Z}_2$ monopole is stable while $\mathbb{Z}_2$ monopoles of indices 2, 4 and 10 are all unstable.

%%%%%%%%%%%%%%%%%%%%%%%%%%%%%%%%%%%%%%%%%%%%%%%%%%%%%%%%%%%%%%%%%%%%%%%%%%%%%%

\section{Discussion on Duality} \label{Duality}

Here we have described a mass hierarchy of spherically symmetric $\mathbb{Z}_2$ monopoles in $SU(4)$ broken to $SO(4)$. The multiplicity of solutions arises due to the variety of ways of embedding $su(2)$ subalgebras into $su(4)$ with a single generator in $so(4)$. Each solution is then labeled by its corresponding embedding index and since indices are gauge invariant, solutions with different indices are physically distinct. Which is further supported by the fact that they display different masses. Due to the topology of the vacuum manifold, particles of higher indices lie on the trivial sector and are expected to decay into stable fundamental particles. Thus we see a hierarchy of massive particles, most of which are unstable, with a nontrivial series of masses.

Now, much like the Goddard-Nuyts-Olive conjecture \cite{GNO1977} concerns $\mathbb{Z}$ monopoles as dual particles to gauge bosons, we consider that $\mathbb{Z}_2$ monopoles may be seen as dual particles to massive fermions in the sense considered in \cite{Strassler1998} where quantum numbers of $\mathbb{Z}_2$ monopoles and massive fermions have been shown to agree. Therefore we propose that masses of fermions of higher generations might be related to masses of dual $\mathbb{Z}_2$ embedded monopoles of higher index. That is to say, the study of $\mathbb{Z}_2$ monopoles might provide a mechanism which yields the Standard Model generations of fermions and their mass hierarchy from a dual gauge theoretical point-of-view.

Note that the fact that higher index monopoles in $su(4)$ all lie in the trivial sector implies fermion number would be violated during decay. To resolve this, we believe a model fulfilling this $\mathbb{Z}_2$ duality hypothesis should support $\mathbb{Z}_2$ monopoles of higher odd index. Table IV in \cite{Lorente1972}, suggests examining embeddings in $su(12)$ of indices 1, 35 and 165 as possible candidates. Secondly, fermion masses are orders of magnitude apart, while $\mathbb{Z}_2$ monopoles in $su(4)$ have comparatively similar masses, as seen in Table \ref{results}. This further motivates the investigation of $\mathbb{Z}_2$ monopoles in larger gauge groups. 

Having these comments in mind, we propose that some model with a large non-Abelian unbroken subgroup satisfying $\pi_2(G/G_0)=\mathbb{Z}_n$ might offer a dual description of fermion generations in the Standard Model. Properties such as their mass hierarchy and perhaps even the Koide formula \cite{Koide}, would be justified by $\mathbb{Z}_n$ soliton particles in a dual theory whose particle content would be organized into a hierarchy like the one described here. 

\section{Conclusion} \label{Conclusion}

In this work we have considered an $SU(4)$ Yang-Mills-Higgs theory spontaneously broken to $SO(4)$ by a scalar field in the symmetric second-rank tensor representation. We explicitly find all the $su(2)$ embeddings for which one of the generators belongs to the unbroken algebra $so(4)$. These embeddings correspond to indices 1, 2, 4, and 10. 

We calculated the branching of the scalar field representation under their respective embedded subgroups so as to decompose the vacuum state into a direct sum of $su(2)$ multiplets. Having this decomposition, we applied the hedgehog transformation to the vacuum state in order to find spherically symmetric solutions in the asymptotic sphere. This gives rise to $\mathbb Z_2$ monopole solutions associated with the previously identified embeddings. Some of the index 2, as well as all index 4 and index 10 solutions are novel results. We found that index 4 monopoles belong to a $su(2)$ quintuplets and index 10 monopoles belong both to $su(2)$ triplets and septuplets. 

Finally we propose radial ansatzes for these solutions and numerically solve their equations of motion in the vanishing potential limit, obtaining profile functions for the scalar and vector fields, as well as the masses and radii of these monopoles in the vanishing potential limit. In particular, index 10 monopoles are found to be approximately twice as heavy as index 4 monopoles. The mass hierarchy observed for indices 1, 2, 4 and 10 monopoles motivated us to explore the idea of a duality between $\mathbb{Z}_2$ monopoles and fermions with the aim of explaining their generations.

\section{Appendix}
The asymptotic field in matrix form may be useful for some calculations, so we provide it below.
Index 1 $\mathbb{Z}_2$ monopole matrix can be obtained by substituting the multiplets in \eqref{asymptotic field} by matrices as in Section \ref{Vacuum Decomposition},
\begin{align}\phi(\theta,\varphi)=\frac{v}{2}
    \begin{pmatrix}
        \cos\theta - i\sin\theta\sin\varphi & \sin\theta\cos\varphi & 0 & 0\\
        \sin\theta\cos\varphi & \cos\theta + i\sin\theta\sin\varphi & 0 & 0\\
        0 & 0 & 1 & 0\\
        0 & 0 & 0 & 1\\
    \end{pmatrix}\label{index 1 scalar field},
\end{align}
This is called the fundamental monopole of magnetic weight $\lambda_1$. Exchanging roots $\alpha_1$ and $\alpha_3$ yields the monopole of weight $\lambda_2$. The resulting scalar field is the permutation of \eqref{index 1 scalar field} in which the embedding occurs in the second diagonal block. Both of these magnetic weights belong to the topological sector $\lambda_1+\Lambda(so(4)^{\vee})$. The asymptotic scalar field of the index 2 $\mathbb{Z}_2$ monopole, of magnetic weight $\lambda_1+\lambda_2$, is expressed as
\begin{align*}
    \phi=\frac{v}{2}\begin{pmatrix}
        \cos\theta - i\sin\theta\sin\varphi & \sin\theta\cos\varphi & 0 & 0\\
        \sin\theta\cos\varphi & \cos\theta + i\sin\theta\sin\varphi & 0 & 0\\
        0 & 0 & \cos\theta - i \sin\theta\sin\varphi & \sin\theta\cos\varphi \\
        0 & 0 & \sin\theta\cos\varphi & \cos\theta + i\sin\theta\sin\varphi \\
    \end{pmatrix},
\end{align*}
This solution is therefore a superposition of two fundamental monopoles of weights $\lambda_1$ and $\lambda_2$. Similarly one can also construct $\lambda_1-\lambda_2$ and $-\lambda_1+\lambda_2$ monopoles. The asymptotic scalar field of a representative index 4 $\mathbb{Z}_2$ monopole is given by the linear combination \eqref{asymptotic field} of the following $T_3$ eigenstates
\begin{align*}
            \ket{2,\pm 2}&=(-e_{11}\mp\sqrt{2}i(e_{12}+e_{21})+e_{13}+e_{31}+2e_{22}\pm\sqrt{2}i(e_{23}+e_{32})-e_{33})/4,\\
            \ket{2,\pm 1}&=(\mp 2ie_{11}+\sqrt{2}(e_{12}+e_{21})+\sqrt{2}(e_{13}+e_{31})\pm 2ie_{33})/4,\\
            \ket{2,0}&=(3e_{11}+(e_{13}+e_{31})+2e_{22}+3e_{33})/\sqrt{24},\\
            \ket{0,0}&=(-(e_{13}+e_{31})+e_{22}+3e_{44})/\sqrt{12}.
\end{align*}
Where we introduce a phase $\ket{2\ 0\ 0}=-e_{11}$ so that $v_b/v$ are real and positive. Finally, the asymptotic scalar field of a representative index 10 $\mathbb{Z}_2$ monopole is composed of a septuplet and a triplet. Using $\ket{2\ 0\ 0}=ie_{11}$ their $T_3$ eigenstates are
\begin{align*}
        \ket{3,\pm 3}&= (\pm ie_{11}-\sqrt{3}(e_{12}+e_{21})\mp i\sqrt{3}(e_{13}+e_{31})+(e_{14}+e_{41})\ +\\
        &\ \ \ \ \ \mp 3ie_{22}\mp 3(e_{23}+e_{32})\pm 3ie_{33}\pm\sqrt{3}(e_{34}+e_{43})\pm ie_{44})/8,\\
        \ket{3,\pm 2}&= (-\sqrt{3}e_{11} \mp 2i(e_{12}+e_{21})+e_{13}+\sqrt{3}e_{22}+e_{24}+e_{42}\ +\\
        &\ \ \ \ \ +\sqrt{3}e_{33}\pm 2i(e_{34}+e_{43})-\sqrt{3}e_{44})/\sqrt{32},\\
        \ket{3,\pm 1}&=(\mp 5i\sqrt{3}e_{11}+5(e_{12}+e_{21})\mp i(e_{14}+e_{41})+\sqrt{3}(e_{13}+e_{31})\ +\\
        &\ \ \ \ \ \mp\sqrt{3}ie_{22}+3\sqrt{3}(e_{23}+e_{32})\mp i(e_{24}+e_{42})\ +\\
        &\ \ \ \ \ \pm\sqrt{3}e_{33}+5(e_{34}+e_{43})\pm5i\sqrt{3}e_{44})/(8\sqrt{5}),\\
        \ket{3, 0}   &= (5e_{11}+\sqrt{3}(e_{13}+e_{31})+3e_{22}+\sqrt{3}(e_{24}+e_{42})+3e_{33}+5e_{44})/(4\sqrt{5}),\\
        \ket{1,\pm 1}&=(\pm\sqrt{3}i(e_{13}+e_{31})-3(e_{14}+e_{41})\mp2ie_{22}+(e_{23}+e_{32})\ +\\ &\ \ \ \ \ \mp\sqrt{3}i(e_{24}+e_{42})\pm2ie_{33})/(2\sqrt{5}),\\ 
        \ket{1, 0}   &= (-\sqrt{3}(e_{13}+e_{31})+2e_{22}-\sqrt{3}(e_{24}+e_{42})+2e_{33})/(2\sqrt{5}).
\end{align*}

    % Bibliography

    \bibliographystyle{JHEP}
    \bibliography{Monopoles.bib}

\end{document}